\documentclass[10pt,twocolumn]{article}
\usepackage[T1]{fontenc}
\usepackage{verbatim}
\usepackage{mathptmx}
\usepackage{listings}
\usepackage{amsfonts}
\usepackage{courier}
\usepackage{geometry}
\usepackage{caption}

\pdfoutput=1
\usepackage[T1]{fontenc}
\usepackage{hyperref}
\usepackage{url}
\usepackage{dcolumn}
\usepackage{graphicx}
\usepackage{color}
\usepackage{geometry}

\usepackage{listings}
\usepackage{color}

\definecolor{mygreen}{rgb}{0,0.6,0}
\definecolor{mygray}{rgb}{0.5,0.5,0.5}
\definecolor{mymauve}{rgb}{0.58,0,0.82}

\usepackage{float}
\newfloat{lstfloat}{htbp}{lop}
\floatname{lstfloat}{Listing}

\begin{document}


\title{\huge{\textbf{Profile Guided Optimization without Profiles:\\*A Machine Learning Approach}}}

\author{Nadav Rotem \\ Meta, Inc. \and Chris Cummins \\ Meta AI}
\date{\today}

\maketitle

\begin{abstract}

Profile guided optimization is an effective technique for improving the optimization ability of compilers based on dynamic behavior, but collecting profile data is expensive, cumbersome, and requires regular updating to remain fresh.

We present a novel statistical approach to inferring branch probabilities that improves
the performance of programs that are compiled without profile guided
optimizations. We perform offline training using information that is collected
from a large corpus of binaries that have branch probabilities information. The
learned model is used by the compiler to predict the branch probabilities of
regular uninstrumented programs, which the compiler can then use to inform optimization decisions.

We integrate our technique directly in LLVM, supplementing the existing human-engineered compiler heuristics. We evaluate our technique on a suite of benchmarks, demonstrating some gains over compiling without profile information. In deployment, our technique requires no profiling runs and has negligible effect on compilation time.

\end{abstract}

\section{Introduction}

Compiler optimizers are structured as a long pipeline, where each stage (pass) in the
pipeline transforms the intermediate representation (IR). Optimization
passes employ different heuristics for making decisions about the optimization
that they perform. For example, the loop unroller has a set of rules for
deciding when and how to unroll loops. The compiler has thousands of heuristics
and threshold parameters that predict profitability of transformations and
balance between performance gains and the cost of compile time.

In this work we replace parts of LLVM's BranchProbabilityInfo (BPI) heuristics,
which have been developed over the last decade by dozens of engineers, with a learned
model.  BPI contains hundreds of rules such as "backedges in loops are hot" and
"branches to terminators are cold" which are coded in C++.  Listing~\ref{fig:rule} presents one of the rules that BPI uses.  We replace these rules
with a pre-computed decision tree that uses inputs that are collected from the
program.

\begin{lstfloat}[t]
\lstset {language=C++}
\begin{lstlisting}[
  caption={An example hand-crafted rule from LLVM's BranchProbabilityInfo (BPI) analysis pass. The BPI analysis contains hundreds of such rules developed over a decade. Our technique replaces these rules with automatically constructed decision trees.},
  label=fig:rule,
]
// Calculate Edge Weights using "Pointer Heuristics".
// Predict a comparison between two pointer or pointer
// and NULL will fail.
bool BranchProbabilityInfo::
     calcPointerHeuristics(const BasicBlock *BB) {
  const BranchInst *BI = dyn_cast<BranchInst>(..);

  Value *Cond = BI->getCondition();
  ICmpInst *CI = dyn_cast<ICmpInst>(Cond);
  if (!CI || !CI->isEquality())
    return false;

 // p != 0   ->   isProb = true
 // p == 0   ->   isProb = false
 // p != q   ->   isProb = true
 // p == q   ->   isProb = false;
 bool prob = CI->getPredicate() == ICmpInst::ICMP_NE;
 ...
\end{lstlisting}
\end{lstfloat}

It is difficult to tune compiler parameters by measuring the performance of the
generated code. System noise and lack of causality are two of the challenges.
To work around this difficulty we formulate a learning task for which deterministic ground-truth labels are readily available. We aim to predict
branch probabilities as they are recorded in the program metadata.  The
mechanism of instrumentation and collection of branch probabilities are already
implemented in the compiler, and we can use them to collect information about
branch behavior. The availability of ground-truth branch probabilities allows us
to turn the problem of replacing BPI into a supervised learning problem.

In this paper we present a new compiler pass that annotates branch probabilities
and does not rely on profile collection. LLVM's BPI uses the branch probabilities
metadata instead of heuristics if the information is available.  Our system
inspects millions of branches from different programs and constructs a
predictive model that can assign branch probabilities to unseen branches.

\begin{figure}[t]
  \includegraphics[width=0.5\textwidth]{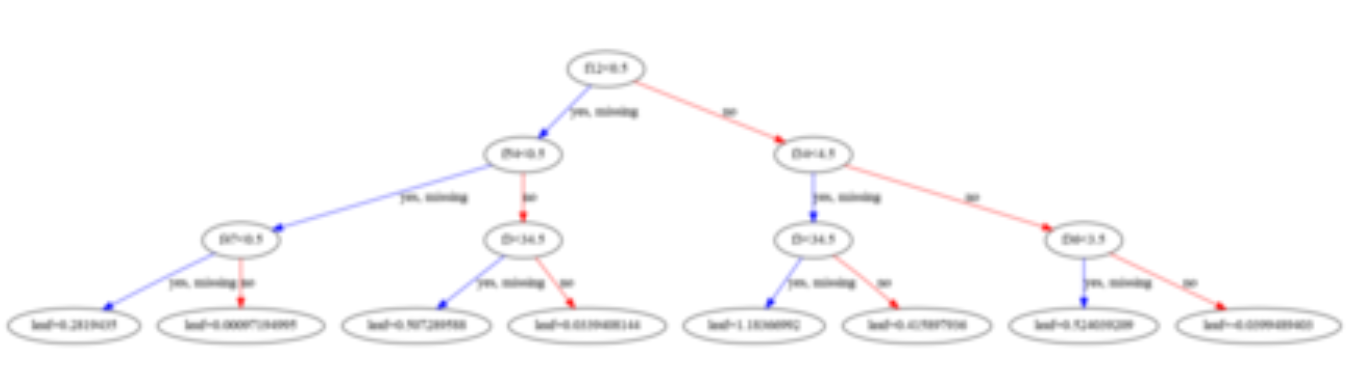}
  \caption{We use profiles collected from training programs to automatically construct decision trees to predict branch weight information for programs for which we do not have profile information. This figure shows one such decision tree. 
  }
  \label{fig:dectree}
\end{figure}

Our focus in this work is to develop a practical solution that can be deployed at scale. For this reason we require that our technique integrate seamlessly into the compiler, add minimal overhead to the compilation pipeline, and be interpretable to aid in development. We chose not to use deep neural networks because of their runtime performance and difficulty of integration. Instead we use
gradient-boosted trees~\cite{boosting} to generate interpretable decision trees.
We use the XGBoost~\cite{xgboost} library for generating decision trees. Decision
trees are similar to the structure of compiler heuristics, except that the rules
and numeric thresholds are automatically computed. Figure~\ref{fig:dectree}
shows a decision tree that was automatically generated by the system.

Unlike LLVM's BPI analysis the rules that our system generates don't need to be
expressed in human language. They can be more complex, can combine more inputs, 
and can use numbers that are not perfectly round.
The code that constructs the decision tree is significantly simpler compared 
to LLVM's BPI which is implemented in thousands of lines of code that 
contain different human readable rules.


We make the following contributions:
\begin{itemize}
    \item We develop a novel statistical approach for predicting branch probabilities from static program features without access to profile information.
    \item We implement our approach for LLVM. We develop a feature extraction pass for LLVM and construct boosted decision tree models from a corpus of training profiles, embedding the constructed model in a new pass for LLVM.
    \item We evaluate our approach on a suite of open source benchmarks. Without access to profile information, and with negligible compilation overhead, we achieve significant performance gains on some programs.
\end{itemize}

\section{Prediction Guided Optimization}

We present a novel technique to statically predict the branch weights of programs without access to profile information. This section describes the design and implementation of the system.

\subsection{Overview of our Approach}

\begin{figure}
  \includegraphics[width=0.5\textwidth]{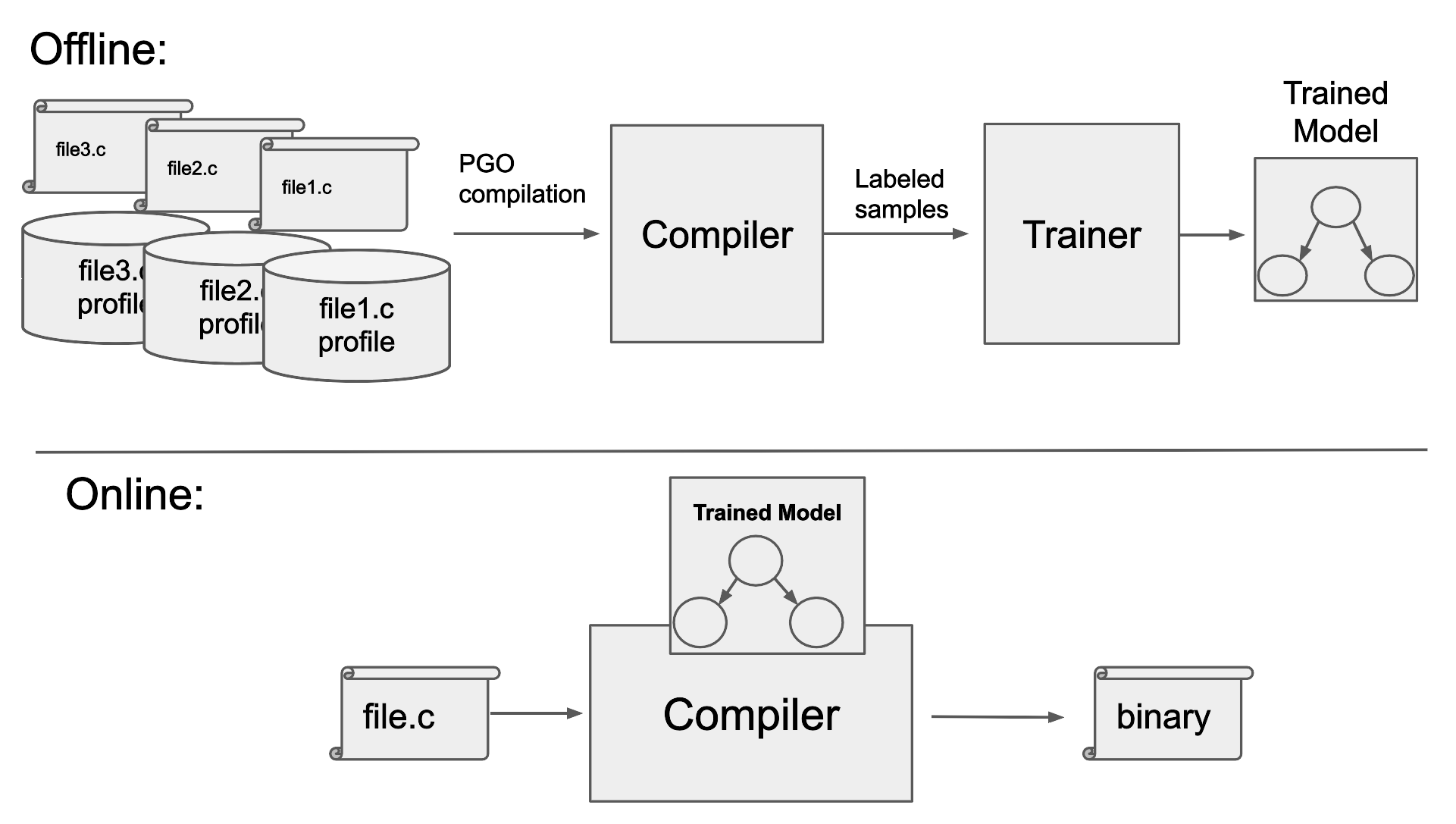}
  \caption{Overview of our approach. During offline training, static features and profiled branch weights are collected from a corpus of training programs and used to automatically construct a model that is then embedded in the compiler. Online, the features of unseen programs are extracted and the model infers branch weights.}
  \label{fig:system}
\end{figure}

Figure~\ref{fig:system} presents the overall compilation flow. The system operates in
two stages: offline training, and online use.

\textbf{Phase I (offline training):} In the first stage a regular compiler uses
the PGO workflow to compile many programs of different kinds. During this
phase a new compiler pass collects information about each branch in the code
(this is the feature list, denoted by X). The compiler also records the branch
probabilities which are provided by the PGO workflow (this is the label, denoted
by Y).  In the context of this discussion we refer to inputs as Features, which
are individual inputs to our prediction engine.  The compiler saves this
information in a large file on disk. Next, an offline script processes the data
and generates a model that can answer the question: given some information about
the branch, what are the most likely branch probabilities? The model is compiled
into C code and integrated into the production compiler.

\textbf{Phase II (regular online compilation):} In this phase the compiler, which is
equipped with a new analysis, compiles regular files without using the PGO
workflow. The new model that was generated in the first phase provides branch
probability information that allows the compiler to make better decisions and
generate faster code.

\subsection{Feature Collection} \label{ref:feature}

In order to generate accurate the branch probabilities our analysis needs to
collect information about how branches behave in different programs. Just like
LLVM's BPI analysis we collects many features using C++ code.  We ask questions
such as, "how many instructions are in each one of the basic blocks that the
branch points to", and "does the right destination block dominate the branch".
We also collect information about the presence of certain instructions such as
return and exception handling instructions. We also record information about the
loop nest, information about the parent block and position within the function.
The pass collects 54 features that we list in \mbox{Appendix \ref{appendix:features}}.
All of the features are stored in a vector of floating point numbers. During the
offline training phase, both the feature-vector and the true branch
probabilities are saved into a file. From this point the features don't have an
assigned meaning and are just numbers.
To ensure a broad sample of the space we collect information from over a
million branches. In our training suite we compile many programs from different
domains: llvm-test-suite benchmarks, abseil, bash, box2d, bzip2, diffutils, distorm, fmt, fpm, graphviz, 
grep, hermes, json-nlohmann, leveldb, liblinear, libpng, libuv, clang, lua, myhtml, oggvorbis, povray, 
python, sela, smallpt, sqlite, tscp, xxhash, z3. We generate representative profile information by
running sample inputs. We ignore branches with too few samples.

\subsection{Training a Model} \label{ref:train}

In the previous stage the compiler saved a large file with millions of rows.
Each row describes the probability of a branch and many features that
describe the branch and the program. We then use standard data
science techniques to generate a model of gradient-boosted trees.

A dataset with n examples and m features is defined as:

\[
 D = \{x_i, y_i\} \; (|D| = n, x_i \in \mathbb{R}^m, y_i \in \mathbb{R}).
\]

A tree-ensemble with K additive functions that predict the label probability:

\[
\widehat{y_i} = \sum_{k=1}^{K} f_k(X_i)
\]

Where each decision tree maps the feature list to a single value.
\[
f : \mathbb{R} ^ m \rightarrow \mathbb{R}
\]

To prepare the data we convert the branch probability values that represent the
ratio between the left-and-right into 11 classes that represent the probabilities $0
.. 1$ in jumps of $0.1$.
We shuffle our data and split it to training and
testing sets using a $10:1$ split. XGBoost is used to generate the decision trees. 
We use the logloss evaluation metric and the softmax learning objective. 
It takes less than a minute to train on data sets with millions of samples.

\subsection{Code generation}

After we finished training our model we need to convert it into C code that can
be included in the compiler.  At inference time, the feature-vector that was
collected for each branch is used by the inference procedure to traverse the
decision tree and reach the desired outcome. Decision trees, such as the one in
Figure \ref{fig:dectree}, turn right or left depending on the values in the
feature list and on the condition at each intersection.

The gradient-boosted trees that we use are an ensemble of weak prediction
models. This means that several simple trees are combined to generate a good
prediction. The probability of each label in the prediction is the accumulation
of several decision trees that process the same input. We convert each tree into
a flat vector and generate a small interpreter that can walk down the tree. At
inference time we process each of the trees and accumulate them into the right
bucket. Finally, we iterate over the 11 labels and find the label with the
highest probability. Figure \ref{fig:tree_c} shows the code that is generated
for a single tree. The generated code can be compiled as part of the LLVM
compiler and serve different passes and analysis.

\begin{lstfloat}[t]
\lstset {language=C}
\begin{lstlisting}[
    caption={The decision tree traversal code generated by our approach.},
    label=fig:tree_c,
]
float intrp(const float *input,
            const short *F, const float *C,
            const short *L, const short *R) {
    int idx = 0;
    while (1) {
      if (F[idx] == -1) return C[idx]; // Found it!
      if (input[F[idx]] < C[idx]) // Check condition.
        idx = L[idx]; // Go left.
      else
        idx = R[idx]; // Go right.
    }
}
// Feature index, condition, left, right:
const short F0[] = {12, 26, 34, 28, 47, 18, 45,  ...
const float C0[] = {0.5, 0.5, 7.5, 1.5, 0.5, 0.5, ...
const short L0[] = {1, 3, 5, 7, 9, 11, 13, 15,17, ...
const short R0[] = {2, 4, 6, 8, 10, 12, 14, 16,   ...

float tree0(const float *input) {
 return intrp(input, F0, C0, L0, R0);
}
\end{lstlisting}
\end{lstfloat}

\subsection{Inference}

This section describes the online use of the compiler. In the system that we've
implemented the only difference to the compiler is a new pass that can assign
branch probabilities based on program structure. During runtime the analysis
collects information on the branches in the program in a feature vector and
passes them to the inference method that was generated off-line. The inference
method returns a value that represents the ratio between the left branch and the
right branch.

\section{Implementation}

Our training system distilled a database of around 200MB into
a model that was around 2MB. The model does not increase the dynamic memory usage of the
compiler because the decision tree tables are stored in read-only memory. The total inference code added to LLVM amounts to less than one kilobyte of code. The
time it takes to infer the properties of a branch depends on the depth of the
decision trees and the number of trees, and there is a tradeoff between accuracy
and performance. However since the decision trees are small and the
traversal is efficient, the overall inference time is very low. The model we trained and evaluate in Section~\ref{eval} can run $250,000$ inference requests per second on a single x86 core. The 3-label model can run $1,600,000$ inferences per second.
This can be further optimized \cite{RapidScorer}.

\begin{figure}
  \includegraphics[width=0.5\textwidth]{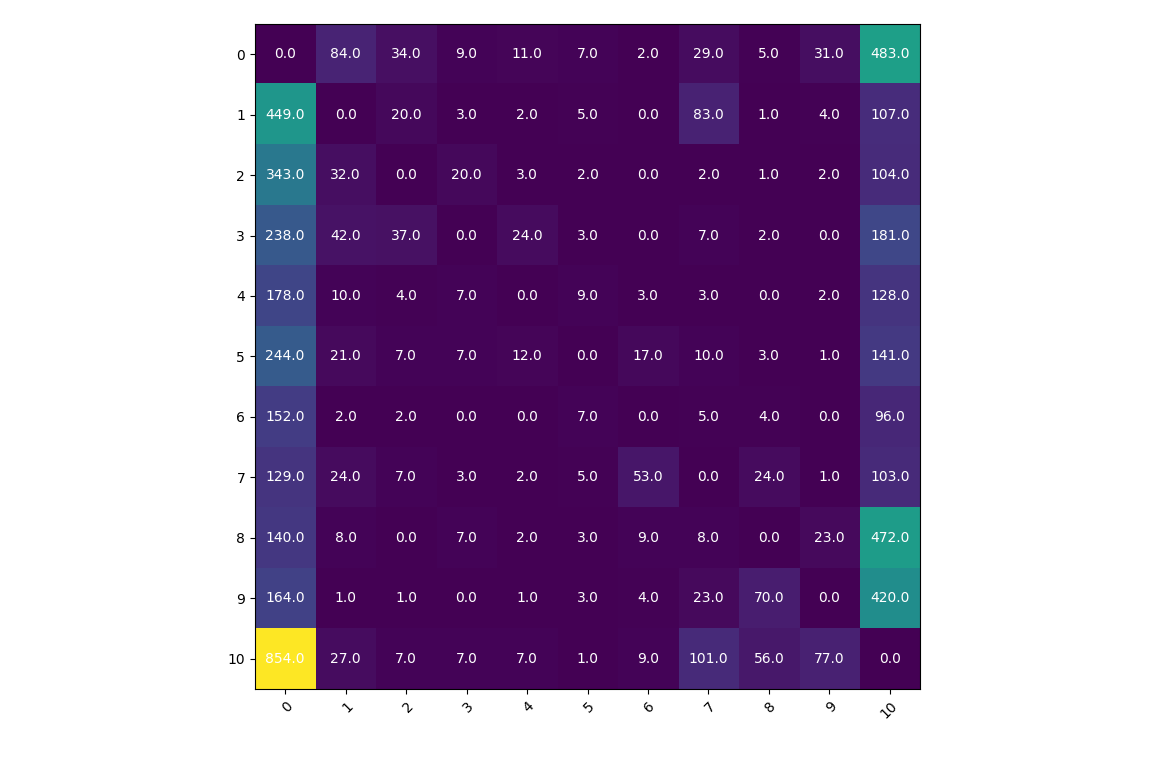}
  \caption{Distribution of branch weight prediction errors. The X axis is ground truth class, the Y axis is predicted class, and cells are brighter with higher density. }
  \label{fig:accuracy}
\end{figure}

\section{ Evaluation }  \label{eval}

We evaluate our technique in two sets of experiments. In the first, we evaluate the predictive performance of our branch weight model. In the second, we evaluate the 
runtime performance of programs compiled using our predicted branch weights and compare it to the performance of programs compiled without any profile information and using ground-truth profiles.

\subsection{Predicting Branch Weights} \label{subsec:predicting-branch-weights}

We use a hold out testing set of 10\% unseen branches to assess whether our model can accurately predict branch weight. We discretize the space of branch weights into 
11 bins: $[0, 0.1, 0.2, \ldots, 1.0]$. Figure~\ref{fig:accuracy} provides an overview of the results. In 75\% of cases the model predicts the correct probability weight class. 
This figure depends on the size and number of the decision trees. 

\subsection{Compilation Performance} \label{subsec:compilation-performance}

\begin{figure}
  \includegraphics[width=0.5\textwidth]{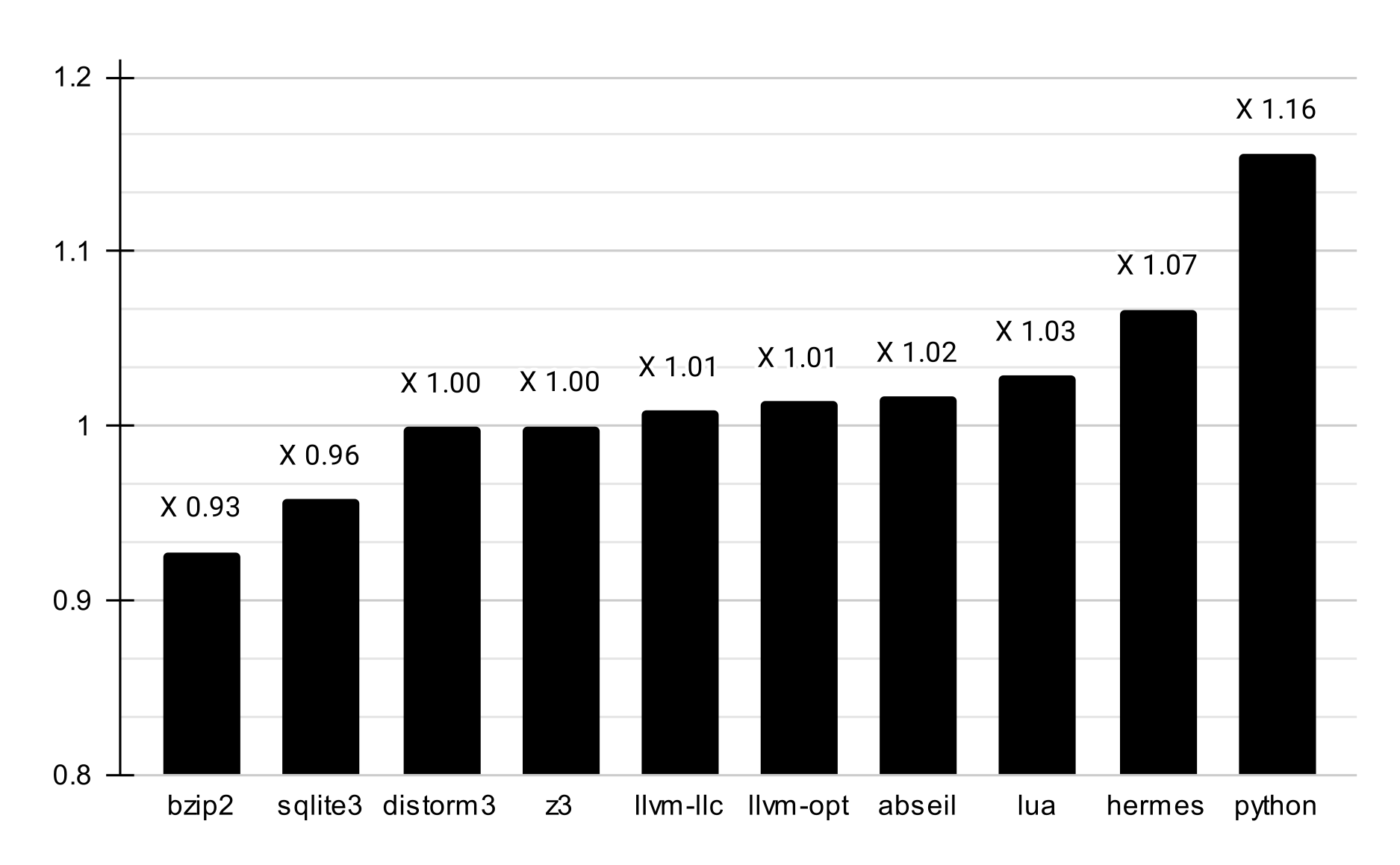}
  \caption{Speedup of programs compiled using our technique over compilation without profile guided optimization. In both cases the compiler does not have access to profile information. The geometric mean speedup of our approach is 1.016.}
  \label{fig:perf}
\end{figure}

We tested our system on a number of different programs from different domains.
Figure \ref{fig:perf} shows the effectiveness of the system on a number of
workloads. In this benchmark we evaluate regular compilation of programs
(without PGO or LTO) using the default optimization compilation flag.
The only difference between the runs was the flag that enabled the new pass. We did not disable BPI, and just allowed BPI to
use the new branch probability metadata. 

The new optimization pass improves the performance of 6 of the 10 workloads, compared to compiling without profile guided optimization. The gains in changes are in the range of $-7\%$ for bzip2 to $+16\%$ for Python.

\subsection{Heuristics Accuracy} \label{subsec:bpi-rate}

In section \ref{subsec:predicting-branch-weights} we evaluate the performance of the machine-learning based approach. In this section we estimate the accuracy of the heuristics in LLVM's BPI analysis. BPI uses four different heuristics when no branch frequency metadata is available. We used the PGO workflow to collect branch probabilities and modified BPI to report both the metadata and the heuristics outcome. We compared the outcome of the "isHot" method on the first successor for each branch. Figure~\ref{fig:bpi_accuracy} shows the effectiveness of each one of BPI's heuristics and the portion of the branches each heuristics handles. 

\begin{figure}
\begin{center}
\begin{tabular}{ |l|l|l|l| } 
 \hline
 Heuristic & Branches & Weight (\%)& Correct(\%)\\
  \hline
Estimated & 2,965,889 & 0.29 & 0.70 \\
Pointer & 4,961,307 & 0.49 & 0.56 \\
Zero & 2,133,775 & 0.21 & 0.76 \\
FloatingPoint & 6,227 & 0.00 & 0.45 \\
  \hline
 \end {tabular}
\end{center} 
 \caption{Accuracy of the heuristics in LLVM's BPI analysis. }
  \label{fig:bpi_accuracy}
\end{figure}


\section{Related work}


Profile guided optimization is a well-established technique for improving compile time optimization decisions~\cite{pgo}. Profile information is
collected using instrumentation or sampling of the executable, and the collected
data can only be used to optimize the executables on which it was trained. Compilers
such as LLVM~\cite{llvm} and BOLT~\cite{BOLT} use profile guided optimizations
to optimize and perform efficient layout of code.

Despite the performance improvements offered by profile guided
optimization, it is cumbersome to integrate into build systems and adds significant overhead. Significant engineering infrastructure is needed to provide deep integration with large evolving codebases~\cite{gprof}.

We propose a novel technique to overcome the limitations of profile guided
optimization using machine learning. The application of machine learning to compiler optimizations is well studied and shows promise in eliminating the human developed heuristics, as surveyed in~\cite{survey_tuning} and~\cite{ml4c}.

The most commonly used technique is supervised learning. Supervised learning has
been applied to a range of problems such as loop
unrolling~\cite{mlunroll,deeptune}, instruction
scheduling~\cite{learning-schedule}, program
partitioning~\cite{learning-partitioning}, heterogeneous device
mapping~\cite{inst2vec,programl}, function inlining~\cite{learning-inlining},
and various optimization heuristics in GCC~\cite{milepost}.

Supervised machine learning algorithms operate on labeled data, but it's not
easy to extract the labeled data from compilers.  In some specific domains, such
as code generation for linear algebra primitives, there is a fixed compilation
pipeline and a program that is reasonable to measure~\cite{tvm}. In traditional
compilers, the size of the optimization space and the complexity of the
optimization pipelines may make it prohibitively expensive to collect training
data, as it requires exhaustively compiling and measuring each program with
every combination of optimization decision to determine the best performance. In
contrast, our approach enables every measurement of a program to be used to
produce ground truth labels for branch probability.

Another challenge is that measurement noise makes it difficult to evaluate the
performance of program. This is especially significant for very small
measurements such as the runtime of individual basic blocks~\cite{ithemal}.
In~\cite{halide-cost-model}, noise in measurements is used a prediction target.
Some projects rely on proxy metrics such as  static analysis of the generated
binaries or code size~\cite{mlgo,compiler_gym}. In contrast to these works, our
approach enables noise-free ground truth measurements to be collected.

A key challenge in supervised learning is feature design. The quality of a
learned model is limited by the quality of the features used. Many prior works
used hand-crafted vectors of numeric
features~\cite{milepost,mlgo,learning-schedule}, however selecting which values
to include in a feature vector is a time consuming and error prone task. An
automatic approach to feature selection was proposed in~\cite{auto-features} but
this requires a cumbersome grammar to be written to describe the space of
features to search over. More recently, deep learning techniques inspired by
natural language modeling~\cite{deeptune,inst2vec,contextual-embeddings} and
graph learning~\cite{programl,cdfg} have been proposed to simplify the task of
feature engineering by automatically inferring high-level features from
low-level input representations, however these techniques sacrifice
interpretability as the inferred latent representations are hard to reason
about. Our approach aims to strike a balance between interpretability and
feature engineering cost by pairing a large number of the readily available
values from LLVM's static analyses with a machine learning algorithm 
that automatically ranks and prunes features.

There is active research around the differentiation of whole
programs~\cite{diffray}, but as of today the whole compilation pipeline is not
differentiable. A different approach is to optimize compilers using
reinforcement learning. Cummins et al.~\cite{compiler_gym} formulate a suite of
compiler optimization problems as environments for reinforcement learning. Ameer
et al.~\cite{mlvec} use reinforcement learning to make vectorization decisions
in the LLVM vectorizer~\cite{vectorizer}. The MLGO project~\cite{mlgo} is a
framework for integrating neural networks into LLVM, targeting the function
inlining heuristic. ESP~\cite{brad97} is an earlier effort to apply machine learning
to the problem of predicting branch probabilities. ESP uses neural networks and
decision trees to predict the likelihood of a branch to be taken.
VESPA~\cite{vespa} extends BOLT and allows the use of machine
learning techniques for predicting branch probabilities. Their approach is similar 
to our work in that they predict branch probabilities that are later used for code
layout and other binary optimizations. Compared to these works which target individual
optimizations at a time, our technique enables a single learned model to benefit
the entire compiler by predicting branch weight metadata that is available to
all optimization passes.

\section{Conclusions}

We investigated the problem of leveraging data science techniques for generating
branch probabilities in uninstrumented programs.  We proposed a fast and simple
system that use gradient-boosted trees.  We tested the proposed system and
demonstrated significant performance wins on several important workloads with
very low compile times and zero additional memory overhead.

The proposed system
is easy to train and integrate and can be the first step in the direction of
applying data science techniques to compiler engineering.  The work presented in
this paper is the result of experimentation in a huge design space. There are
opportunities for improvement on top of the existing work in every stage, and
there are many tradeoffs that need to be explored.

\appendix

\section{Feature list} \label{appendix:features}
This section lists the features that are extracted from each branch and from the current, left and right basic blocks. The feature extraction code is written in C++ that converts them into a vector of floating point numbers.

\begin{center}
\begin{tabular}{ |l|l| } 
 \hline
 Type & List of features  \\ 
  \hline
 Branch Features  &is\_entry\_block\\
   &num\_blocks\_in\_fn \\
   &condition\_cmp \\
   &condition\_predicate \\
   &condition\_in\_block \\
   &predicate\_is\_eq \\
   &predicate\_is\_fp \\
   &cmp\_to\_const \\
    &left\_self\_edge\\
    &right\_self\_edge\\
    &left\_is\_backedge\\
    &right\_is\_backedge\\
    &right\_points\_to\_left\\
    &left\_points\_to\_right\\
    &loop\_depth\\
    &is\_loop\_header\\
    &is\_left\_exiting\\
    &is\_right\_exiting\\
    &dominates\_left\\
    &dominates\_right\\
    &dominated\_by\_left \\
    &dominated\_by\_right\\
    &num\_blocks\_dominated\\
  \hline 
Basic Block Features &num\_instr\\
    &num\_phis\\
    &num\_calls\\
    &num\_loads\\
    &num\_stores\\
    &num\_preds\\
    &num\_succ\\
    &ends\_with\_unreachable\\
    &ends\_with\_return\\
    &ends\_with\_cond\_branch\\
    &ends\_with\_branch\\
 \hline
\end{tabular}
\end{center}

\bibliographystyle{unsrt}
\bibliography{paper.bib}

\end{document}